\documentclass[10pt]{iopart}
\usepackage{epsfig,cite}
\newcommand{\ewxy}[2]{\setlength{\epsfxsize}{#2}\epsfbox[30 30 640 640]{#1}}
\newcommand{\beq}{\begin{equation}}
\newcommand{\eeq}{\end{equation}}

\newcommand{\beqar}{\begin{eqnarray}}
\newcommand{\eeqar}{\end{eqnarray}}

\begin{document}

\title{ Freeze-out of strange hadrons at RHIC }
\author{
L~V~Bravina\dag\ddag, K~Tywoniuk\dag, E~E~Zabrodin\dag\ddag
}
\address{\dag\
         Department of Physics, University of Oslo, Oslo, Norway}
\address{\ddag\
         Institute for Nuclear Physics, Moscow State University, 
         Moscow, Russia}

\begin{abstract}
The production and freeze-out conditions of strange particles, 
produced in Au+Au collisions at RHIC energies, are studied within 
microscopic transport model. The system of final particles can be 
represented as a core, containing the particles which are still 
interacting with each other, and a halo, in which the particles 
have already decoupled from the system. In microscopic calculations 
hadrons are continuously emitted from the whole reaction volume. At 
RHIC, however, significant fractions of both mesons and baryons are 
emitted from the surface region within the first two fm/c. Different 
species decouple at different times. Strange mesons (kaons and 
$\phi$) are frozen at earlier times and, therefore, can probe 
earlier stages of the reaction. 
\end{abstract}

\section{Space-time freeze-out picture}
\label{sec1}

The main goal of the experiments with ultrarelativistic heavy-ion
collisions at Brookhaven and CERN is to study properties of hot and 
dense hadronic matter. Using the measured final-state distributions 
one tries to reconstruct the dynamical picture of the nuclear 
reaction and compare it with the predictions of different models.
The reaction dynamics and experimental constraints are so complicated 
at high energies that any single model cannot fully reproduce the 
whole set of data. But the discrepancies between the standard models
and the data may help to reveal new phenomena associated with highly 
anticipated quark-hadron phase transition.

In the present paper we continue to study the freeze-out conditions
in relativistic heavy ion collisions which were initiated earlier
\cite{fr_ags,fr_sps}.
The calculations are carried out within the quark-gluon string model
(QGSM) \cite{qgsm} for gold-gold collisions at $\sqrt{s} = 130$ AGeV.
The phase-space distribution for the particles on the mass shell is a
function of seven independent variables: components of radius 
${\vec r}$ and momentum ${\vec p}$, and time $t$. For the sake of 
simplicity we integrate it over some variables and study separately 
different space-time and phase-space three-dimensional distributions.
Note, that production and freeze-out of strange hadrons at RHIC
energies have been studied also in \cite{bleicher} within the UrQMD 
model.

\begin{figure}[htb]
\hspace{2.0cm}
\ewxy{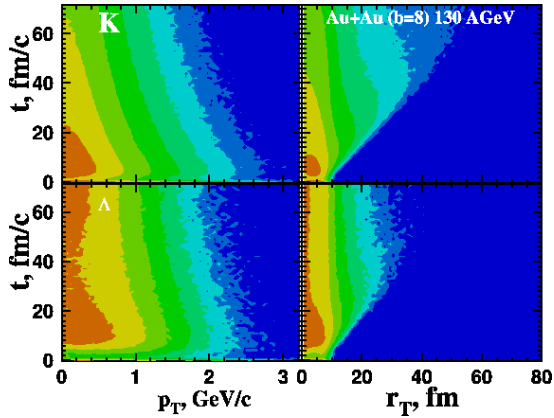}{80mm}  
\vspace{-1.5cm}
\caption{
$d^2N/p_Tdp_Tdt/A$ and $d^2N/r_Tdr_Tdt/A$ distributions of the
final-state $K$ and $\Lambda$ over their last elastic and inelastic 
collision points in the $(p_T,t)$ and $(r_T,t)$ planes.
Particles are produced in Au+Au collisions with $b = 8$~fm at 
$\sqrt{s} = 130$~AGeV. Contour plots 
correspond to different particle densities.
}
\label{fig1}
\end{figure}

Figure~\ref{fig1} shows the distributions $d^2N/p_Tdp_Tdt$ and
$d^2N/r_Tdr_Tdt$ of the emitted kaons and lambdas over time $t$ and 
transverse radius and momentum, $r_T=\sqrt{(x^2+y^2)}$ and 
$p_T=\sqrt{(p_x^2+p_y^2)}$, respectively. 
It is different for $\Lambda$ and $K$. 
One can see in Fig.~\ref{fig1} (right panels) that, in contrast to SPS
\cite{fr_sps}, not only pions but also many kaons
and lambdas are emitted from the surface region $r_T \approx R_A$
within first few fm/$c$. A strong collective transverse expansion of
hadronic matter is observed. Kaons with large transverse momenta are 
emitted predominantly at the initial stages of the reaction 
(Fig.~\ref{fig1}, left panels). They are produced in inelastic primary 
$NN$ collisions, whereas soft hadrons are emitted during the whole
evolution time. At later times the transverse momenta are generated
to a large extent by multiple rescattering. Both due to formation
time effect and longer mean free path of kaons, similar distribution 
for lambdas is rather flat. This plateau corresponds to the
``thermal" component of the $\Lambda$ distribution due to many elastic
and inelastic collisions. Still, the lambdas with maximum
$p_T$ are produced at the beginning of the collision. With growing
time the transverse momentum spectra become gradually softer, which 
can be interpreted as the cooling of the expanding hadronic matter.

\section{Sequential freeze-out}
\label{sec2}

\begin{figure}[htb]
\vspace{-2.0cm}
\hspace{1.0cm}
\ewxy{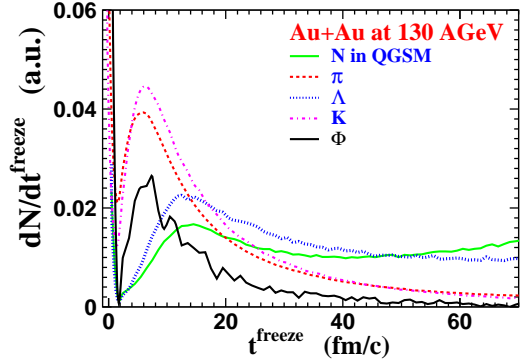}{85mm}  
\caption{
$dN/dt$ distribution of $N, \pi, K, \Lambda$ and $\phi$ over
their last collision time $t^{freeze}$. 
}
\label{fig2}
\end{figure}

Let us consider in detail the time distributions for the
different hadron species shown in Fig.~\ref{fig2}. 
First of all,
there is a noticeable difference between the meson and baryon groups
of particles. The QGSM predicts that kaons and pions decouple earlier
than nucleons and lambdas and approximately at the same times
$\langle t^{\rm mes} \rangle \approx 6$ fm/$c$ and $\langle
t^{\rm bar} \rangle \approx 13$ fm/$c$.
The width of $dN/dt$ distributions for mesons are narrower
than that for baryons: $\Delta t^{\rm mes}\approx 8$ fm/$c$
and $\Delta t^{\rm bar}\approx 14$ fm/$c$.
For the $K$'s and $\Lambda$'s the width is slightly smaller than the
width for pions and nucleons, respectively.
At the last stages of the reaction the $dN/dt$ distributions
for nucleons and pions are determined mainly by the resonance decays
$\Delta \rightarrow \pi + N$, while the width of the distributions
of kaons and lambdas is determined by the elastic collisions.
At this stage $K$ and $\Lambda$ (as well as $\pi$ and $N$) have
the same decoupling times and the slopes of $dN/dt$ distributions.

Therefore, the microscopic model calculations show that there is no  
sharp freeze-out of particles at RHIC. In fact, the particles are 
emitted continuously. In contrast to assumptions of ideal hydrodynamic 
model \cite{Lan53}, the expanding fireball in microscopic models can 
be rather treated as a core consisting of still interacting hadrons, 
and a halo, which contains particles already decoupled from the 
system. The order of the freeze-out of different species seems to be 
the same for energies ranging from AGS to RHIC: 1 - pions, 2 - kaons, 
3 - lambdas, 4 - nucleons. This conclusion is valid also for the 
particles emitted in a certain rapidity interval, particularly at 
central rapidities, $|y| \leq 1$.
On the other hand, the bulk production of mesons takes place within 
8-10 fm/$c$, and the scenario of at least sequential freeze-out of 
hadrons is not ruled out.

Results of our microscopic calculations justify the implementation 
of the continuous freeze-out of particles in a more realistic 3+1 
dimensional hydrodynamic model in \cite{SGHK04}.
It appears, for instance, that this approach can be accounted for
better description of experimental data on the two-particle 
interferometry at RHIC. 

\section{Freeze-out and elliptic flow}
\label{sec3}

The continuous freeze-out of particles modifies distributions
connected to asymmetry of the system. One of this signals which
is intensively studied now is elliptic flow, $v_2$ \cite{Olli92}. 
\begin{figure}[htb]
\begin{minipage}[t]{65mm}
\vspace{-1.0cm}
\hspace{-0.cm}
\ewxy{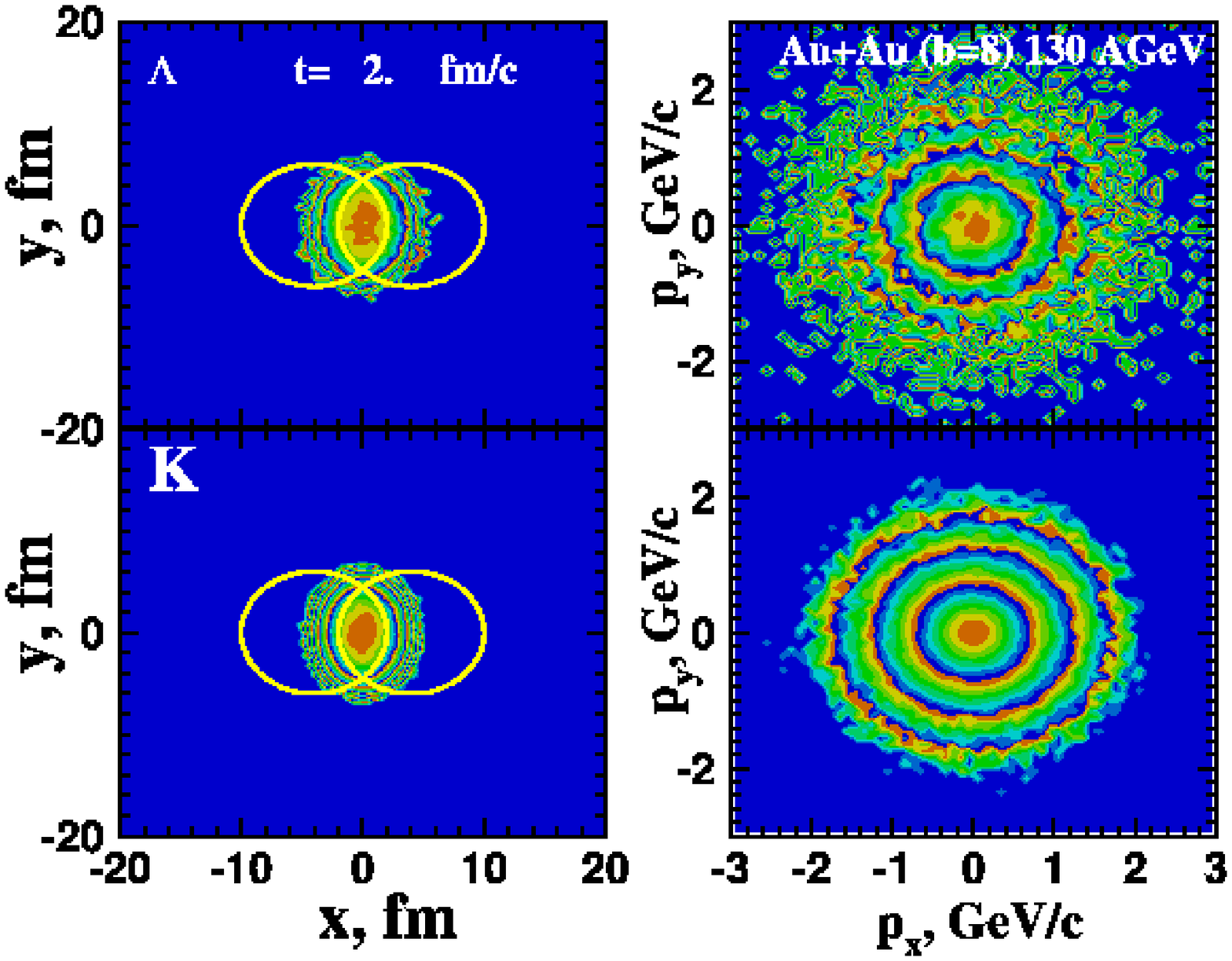}{75mm}  
\vspace{-1.0cm}
\caption{
$d^2N/dx dy$ distributions for $\Lambda$ and $K$ in Au+Au 
collisions with $b = 8$ fm at $\sqrt{s} = 130$ AGeV.
}
\label{fig3}
\end{minipage}
\hspace{\fill}
\begin{minipage}[t]{65mm}
\vspace{-1.0cm}
\hspace{0.5cm}
\ewxy{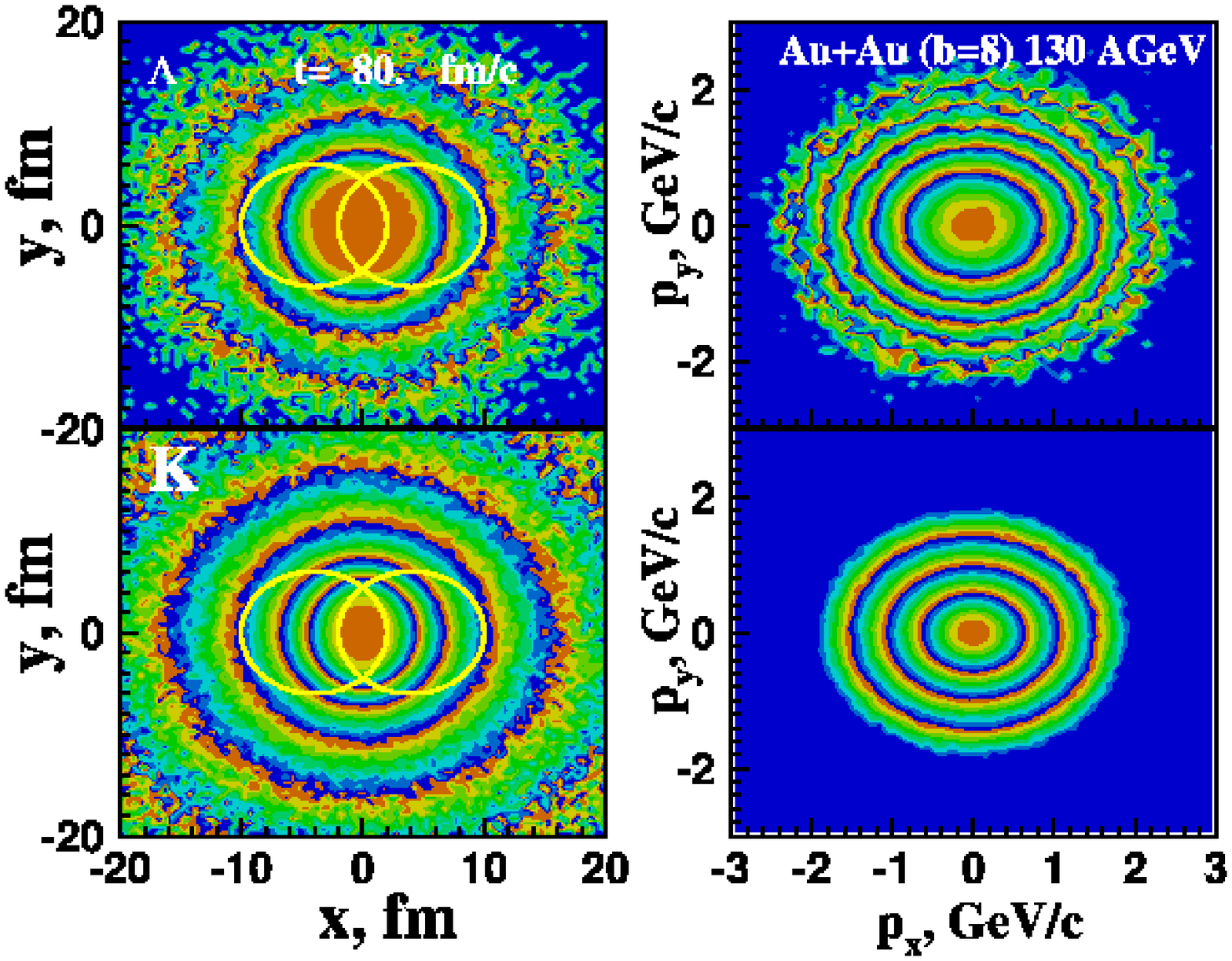}{75mm}  
\vspace{-1.0cm}
\caption{
$d^2N/dp_x dp_y$ distributions for $\Lambda$ and $K$ in Au+Au 
collisions with $b = 8$ fm at $\sqrt{s} = 130$ AGeV.
}
\label{fig4}
\end{minipage}
\end{figure}
Figures~\ref{fig3} and \ref{fig4} depict anisotropy of distributions
for kaons and lambdas in the transverse coordinate $(x,y)$ and 
momentum $(p_x,p_y)$ planes at the beginning and at the end of the
collision, respectively. At the initial stage of the reaction the
distributions of particles possess strong anisotropy in the coordinate
space, but quite weak anisotropy in the momentum one. Then the
momentum-space anisotropy starts to develop for low-momentum 
particles, and at the final stage both kaons and lambdas show quite
noticeable momentum anisotropy, i.e. elliptic flow. 
It is worth mentioning that the final distribution $d^2N/dx dy$ for 
$\Lambda$ has wide plateau located between the centers of colliding
nuclei. For kaons this distribution is much narrower; like pions,
kaons are mostly concentrated within the overlapping region.
Our analysis shows \cite{elfl_04} that the baryonic and mesonic 
components of the elliptic flow are
completely different: pions and kaons emitted from the surface of 
the expanding fireball within the first few fm/$c$ carry the strongest 
flow, while later on their flow is significantly reduced. In contrast 
to this, the baryon fraction acquires stronger elliptic flow during 
the subsequent rescatterings, thus developing the hydro-like flow.
The saturation of the flow at the late stages can be explained
by the lack of rescattering, since the expanding system becomes
more dilute. Further details
of the evolution of $v_2$ are considered in Ref.~\cite{za_04}.

\section{Conclusions}
\label{concl}

Our main conclusion is that the quark gluon string model predicts a
continuous emission of particles, starting almost from the beginning
of the reaction. This can be attributed to the lack of attractive 
forces which keep the particles together. To get a sharp freeze-out it 
is necessary to have some glue mechanism, like attractive mean fields, 
enhanced cross sections or rapidly hadronizing quark-gluon plasma. 
Different species decouple at different times.
The order of the freeze-out of hadronic species is as
follows: 1 - pions; 2 - kaons; 3 - lambdas; 4 - nucleons.
At RHIC energies significant fractions of mesons and baryons
are emitted within the first two fm/$c$.
The mesons with large  transverse
momenta $p_t$ are predominantly produced at the early stages of the
reaction. The low $p_t$ component is populated by mesons coming mainly
from the decay of resonances. This explains naturally the decreasing
source sizes with increasing $p_t$, observed in Hanbury-Brown$-$Twiss
interferometry.

\end{document}